# Grain Refinement and Enhancement of Critical Current Density in the $V_{0.60}Ti_{0.40}$ Alloy Superconductor with Gd addition

Sabyasachi Paul,[1,2] SK Ramjan,[1,2] R Venkatesh,[3] L S Sharath Chandra,[1] M K Chattopadhyay[1,2]

[1]Free Electron Laser Utilization Laboratory, Raja Ramanna Centre for Advanced Technology, Indore 452013, India;
[2]Homi Bhabha National Institute, Training School Complex, Anushakti Nagar, Mumbai 400094, India;
[3]UGC-DAE Consortium for Scientific Research, Takshashila Campus, Khandwa Road, Indore 452017, India

*Email:paulrana11@gmail.com*

*Abstract*— The V-Ti alloys are promising materials as alternate to the commercial Nb-based superconductors for high current-high magnetic field applications. However, the critical current density ($J_c$) of these alloys are somewhat low due to their low grain-boundary density. We show here that grain refinement of the V-Ti alloys and enhancement of the $J_c$ can be achieved by the addition of Gd into the system, which precipitates as clusters along the grain boundaries. Both the $J_c$ and the pinning force density ($F_P$) increase with the increasing Gd content up to 1 at. % Gd, where they are more than 20 times higher than those of the parent $V_{0.60}Ti_{0.40}$ alloy. Introduction of Gd into the system also leads to ferromagnetic (FM) correlations, and the alloys containing more than 0.5 at. % Gd exhibit spontaneous magnetization. In spite of the FM correlations, the superconducting transition temperature increases slightly with Gd-addition.

## I. INTRODUCTION

THE V-Ti alloys are considered to be promising materials as an alternate to Nb-based superconductors for high field applications and in the neutron radiation environment [1]. They are machinable and have high upper critical field ($H_{c2}(0)$) [1, 2]. Our previous studies reveal that the grain boundaries pin the flux lines in these alloys and due to very large grain sizes, the critical current density ($J_c$) is about two order of magnitude lower than the commercial superconductors [3, 4]. The average grain size in the V-Ti alloys vary from few hundred $\mu$m to few mm [4], whereas in the high-$J_c$ commercial $Nb_3Sn$ and Nb-Ti superconductors, the grain size vary between tens to few hundreds of nm [5, 6]. Thus, reducing the grain size may enhance the $J_c$ of the V-Ti alloys. The rare earth (RE) elements have been extensively used for grain refinement in various steel and high entropy alloys, where the RE elements mostly get segregated along the grain boundaries [7-10]. The RE elements being immiscible in the *bcc* phase [11], introducing RE elements into the *bcc* V-Ti alloys may be effective in grain refinement and enhancing the critical current density. It is also known that magnetic inclusions are more effective in pinning the flux lines as compared to normal metallic inclusions or normal point defects [12]. In this direction, we have added gadolinium (Gd) to $V_{0.60}Ti_{0.40}$ alloy and studied its effect on the grain size and $J_c$.

We observe that the grain size reduces with increasing x, which results in the enhancement of the $J_c$ and $F_P$ by more than an order of magnitude in magnetic fields up to 6 T for the alloy with 1 at. % Gd. The addition of Gd into the system also introduces ferromagnetic (FM) correlations. Even in the presence of ferromagnetism a robust superconducting state persists in these alloys, and in fact the superconducting transition temperature increases slightly after Gd addition.

## II. EXPERIMENTAL DETAILS

The polycrystalline $V_{0.60-x}Ti_{0.40}Gd_x$ (x = 0, 0.005, 0.01, 0.02) were prepared by melting the constituent elements (purity > 99.8%) in an arc melting furnace under 99.999% pure Ar-atmosphere. The samples were flipped and re-melted 5 more times to improve the homogeneity. The details of sample preparation for metallograpy are already described in ref [13]. The Scanning Electron Microscopy images were taken using a NoVa NanoSEM apparatus (FEI Company, USA). The compositional analysis of the samples was also performed simultaneously using a X-Flash 6130 energy dispersive spectroscopy attachment (Bruker, Germany) and Esprit software. The temperature and field dependence of magnetization was measured in 7T SQUID-based magnetic properties measurement system (MPMS, Quantum Design, USA).

## III. RESULTS AND DISCUSSIONS

Figure 1 shows the metallographic images of $V_{0.60-x}Ti_{0.40}Gd_x$ alloys. The regions similar to that marked by '$\Delta$' contain only V and Ti, whereas Gd is found only in the clusters (marked by the arrow). The alignment of Gd in closed pattern (shown by dashed lines) indicates that they precipitate along the grain boundaries. The grain size of the $V_{0.60}Ti_{0.40}$ alloy is of the order of few hundred microns which reduces to few tens of microns for the x = 0.02 alloy.

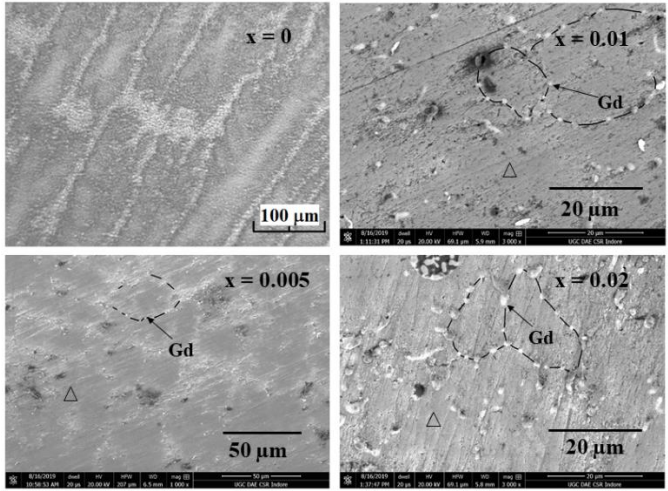

Fig. 1. Metallography images of the $V_{0.60-x}Ti_{0.40}Gd_x$ alloys. The regions similar to that marked by 'Δ' contain only V and Ti. The grain size reduces with Gd addition, and Gd precipitates in clusters of size < 1.2 μm.

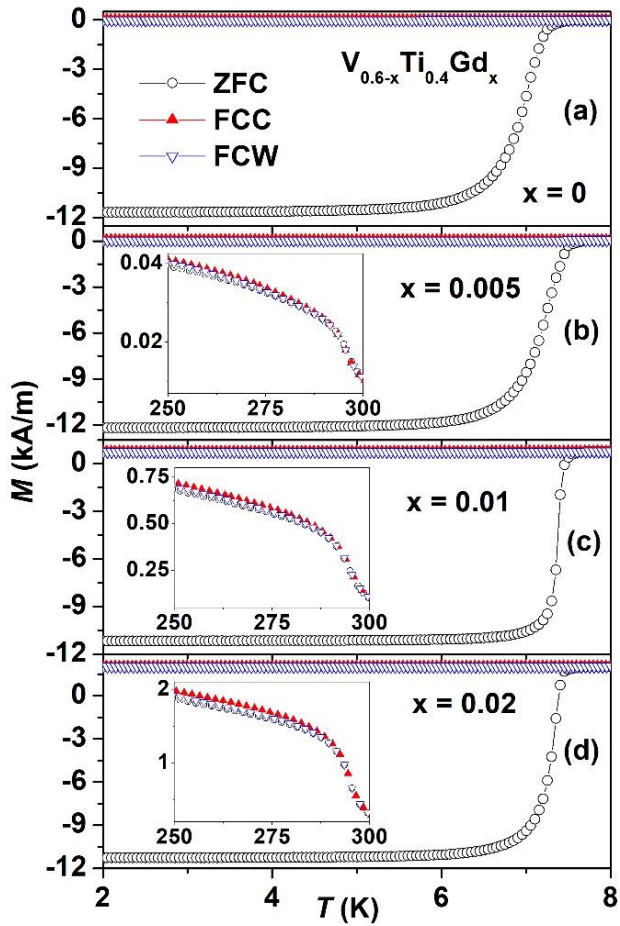

Fig. 2(a)– (d). Temperature dependence of the magnetization of $V_{0.60-x}Ti_{0.40}Gd_x$ alloys measured in the presence of 10 mT magnetic field. The superconducting transition temperature increases slightly with the addition of Gd. Inset to fig. 2(b)– (d) shows a change of slope at 295 K indicating paramagnetic to ferromagnetic phase transition.

Figure 2(a)- (d) shows the temperature dependence of magnetization ($M(T)$) in the temperature range 2– 8 K for an applied field $H$ = 10 mT. The $M(T)$ was measured during the warming up of the sample in $H$ = 10 mT after cooling in zero field (ZFC measurement), cooling in the same field after the ZFC cycle (FCC measurement) and warming up in the same field after FCC cycle (FCW measurement). The sharp drop in $M(T)$ below 7.55 K indicates the normal to superconducting phase transition for the x = 0 alloy. The superconducting transition temperature ($T_{sc}$) increases slightly with the addition of Gd. Large difference in $M(T)$ between the ZFC and FCC measurements indicates strong flux pinning in these alloys. The inset to fig. 2(a)– (d) shows the $M(T)$ in the presence of 10 mT field in temperature range 250 – 300 K. A change of slope in $M(T)$ similar to a paramagnetic to ferromagnetic transition is observed around 295 K (which corresponds to elemental Gd [14]) for all the alloys with x > 0.

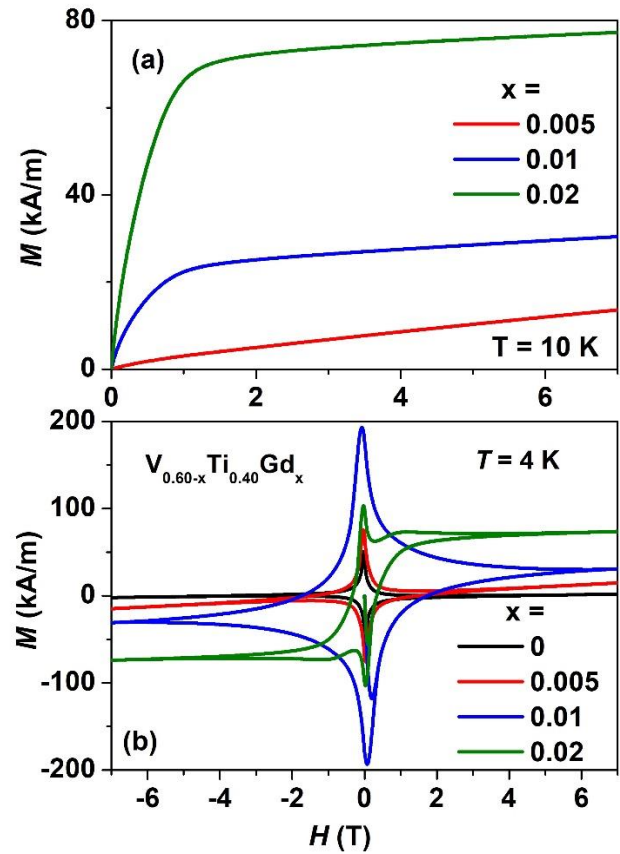

Fig. 3. Field dependence of magnetization ($M(H)$) at (a) 10 K and (b) 4 K. The $M(H)$ at 10 K shows saturation of magnetization for the x > 0.005 alloys. The field hysteresis at 4 K increases with the addition of Gd.

Figure 3 shows the field dependence of magnetization ($M(H)$) of the $V_{0.60-x}Ti_{0.40}Gd_x$ alloys for (a) 10 K ($T > T_{sc}$) and (b) 4 K ($T < T_{sc}$). The $M(H)$ at 10 K shows saturation of magnetization for the alloys with x > 0.005. The Arrott plots [15] reveal small spontaneous magnetization (0.04 and 0.13 $\mu_B$/f.u. respectively for x = 0.01 and 0.02) at 10 K. The absence of any hysteresis at 10 K shows that the ferromagnetism is soft in nature. The hysteresis between increasing and decreasing magnetic fields ($\Delta M$) below $T_{sc}$ increases significantly with increasing x up to x = 0.01. With increasing x, the asymmetry in $M(H)$ also increases. For the x = 0.02 alloy at 4 K, the $M(H)$ shows a sharp peak around $H = 0$ and is almost constant for $|H| > 2$ T. These results indicate that the superconductivity and ferromagnetism coexist in these alloys.

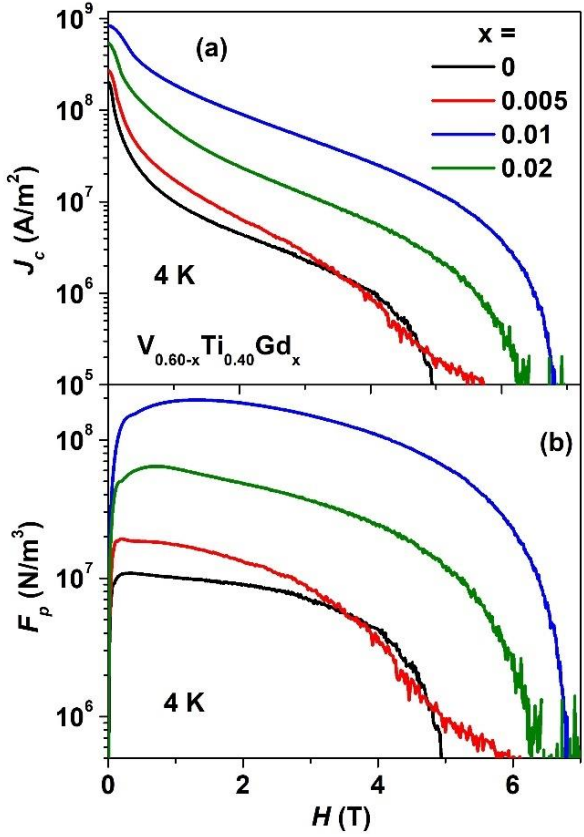

Fig. 4. Field dependence of (a) critical current density ($J_c$) and (b) flux pinning force density ($F_P$) of the $V_{0.60-x}Ti_{0.40}Gd_x$ alloys. Both $J_c$ and $F_P$ increases with the addition of Gd and the enhancement is maximum for the alloy with 1 at. % Gd.

The critical current density ($J_c$) of the present alloys are estimated using the Bean's critical state model [16]. For a rectangular sample, the $J_c$ is given by [3, 17],

$$J_c = \frac{2\Delta M}{a\left(1 - a/3b\right)}, \quad (1)$$

where $a$ and $b$ ($a < b$) are respectively the width and thickness of the sample perpendicular to the direction of the applied magnetic field. The pinning force density ($F_P$) has been estimated using the formula [18], $F_P = J_c \times H$. Figure 4 (a) and (b) respectively shows the $J_c$ and $F_P$ of the $V_{0.60-x}Ti_{0.40}Gd_x$ alloys at 4 K. Both the $J_c$ and $F_P$ increase with the addition of Gd in the $V_{0.60}Ti_{0.40}$ alloy. The enhancement is maximum for the x = 0.01 alloy where it is more than 20 times higher than the parent alloy. The $J_c$ of the alloy with 1 at. % Gd is now only an order of magnitude lower than the commercial NbTi superconductors. The irreversibility field ($H_{irrv}$) (applied magnetic field value for which $J_c$ becomes zero), also increases with the addition of Gd. The $J_c$, however is found to reduce for the x = 0.02 alloy probably due to the presence of stronger ferromagnetic correlations.

The field dependence of pinning force density ($F_P(H)$) behavior may be understood using the Dew-Hughes model [19], where the normalized pinning force density $f_P (= F_P/(F_P)_{max})$ can be approximated as, $f_P \propto h^p(1-h)^q$, where $h = H/H_{irrv}$ is the reduced field and the value of $p$ and $q$ depend on the type of the pinning mechanism [18, 19]. The functional form of $f_P(h)$ corresponding to different pinning mechanisms are listed in Table 1.

Table 1: Details of $f_P(h)$ according to Dew-Hughes model [19]

| Curve no. in Fig. 5 (a) | Geometry of pin | Type of pinning centre | Functional form | Position of the peak |
|---|---|---|---|---|
| 1 | Volume | $\Delta\kappa$ | $h(1-h)$ | 0.5 |
| 2 | Surface | Normal | $h^{0.5}(1-h)^2$ | 0.2 |
| 3 | Surface | $\Delta\kappa$ | $h^{1.5}(1-h)$ | 0.6 |
| 4 | Point | Normal | $h(1-h)^2$ | 0.33 |
| 5 | Point | $\Delta\kappa$ | $h^2(1-h)$ | 0.67 |

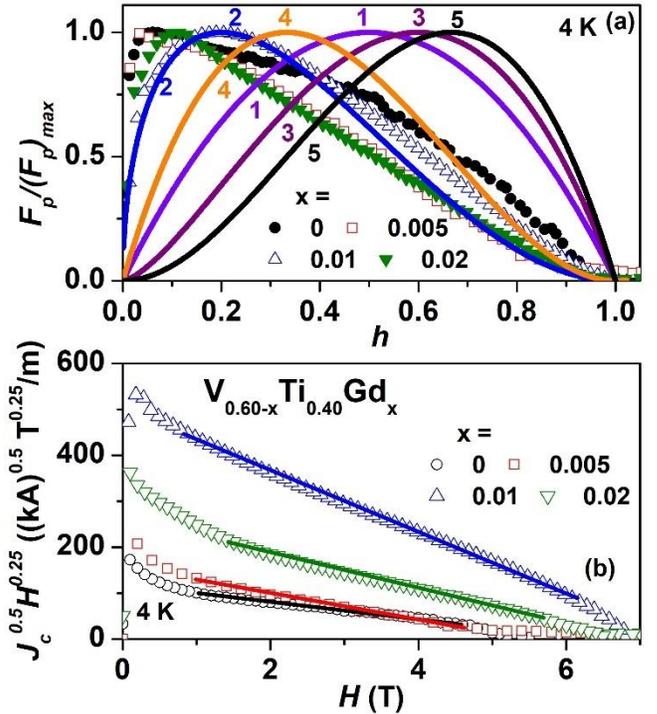

Fig. 5 (a). Normalized pinning force density as a function of $h = H/H_{irrv}$ at 4 K. The solid lines represent different functions shown in table 1. (b) Kramer plots for the alloys at 4 K, where the increase in the linear region indicates that pinning at grain boundaries increase with Gd-addition.

The $f_P(h)$ curves exhibit a single peak structure when only one pinning mechanism dominates [19]. Thus, the multiple shoulder structure of the curves of Fig. 4(b) indicate the presence of more than one flux line pinning mechanisms. Figure 5(a). depicts more clearly the presence of multiple pinning mechanisms in these alloys. In this latter figure, the experimental $f_P(h)$ curves (data points) of the present alloys at 4 K are plotted along with the relevant theoretical [19] lines explained in Table 1. In the table, $\kappa$ is the Ginzberg-Landau parameter and $\Delta\kappa$ refer to the change of the superconducting properties due to metallic pinning centres influenced by proximity effect of the superconducting matrix [19]. The theoretical lines 2 and 3 in figure 5 (a) respectively represent the flux pinning at the grain boundaries and dislocations [4, 19]. From figure 5 (a), we can observe that the point defects and dislocations play significant role to pin the flux lines in the $V_{0.60}Ti_{0.40}$ alloy. However, this changes with the addition of Gd and for the alloy with 1 at. % Gd, the grain boundaries are the major pinning centers. This is consistent with the Kramer's model, according to which, the $J_c^{1/q} H^{(1-p)/q}$ vs. $H$ plots approximates to straight lines over the magnetic field range where the corresponding pinning mechanism is dominant [20]. The $J_c^{0.5} H^{0.25}$ vs $H$ plot (corresponding to $p = 0.5$ and $q = 2$) is known as Kramer plot and indicates the field regime over which the pinning at the grain boundaries dominate. The linearity in the Kramer plot for the x = 0 alloy is observed from 1- 4.6 T magnetic fields. The field regime over which the Kramer plot shows linearity increases with Gd addition and for x = 0.01, the Kramer plot is linear from 0.8- 6.2 T. This shows that the field regime over which the flux line pinning due to the grain boundaries dominates, increases with the addition of Gd. This is also consistent with our metallography results, where we observe an increase in the grain boundary density (smaller grains) with increasing Gd content. As both the size and density of the Gd precipitates increase with increasing x, the flux line pinning mechanisms of the x = 0.02 alloy is probably affected by the stronger FM correlations. We recall here that our previous studies show that the V-Ti alloys have itinerant spin fluctuations that suppress the $T_{sc}$ to about half its theoretical value [21]. The FM clusters of Gd mentioned above might affect the coherence of the spin fluctuations, and this might be the reason for the slight enhancement of the $T_{sc}$ with Gd addition.

## IV. Conclusion

The present results show that the addition of Gd in the $V_{0.60}Ti_{0.40}$ alloy is effective in grain refinement, which leads to the enhancement of $J_c$. The $J_c$ and $F_P$ are the maximum for the alloy with 1 at. % Gd, where they are more than 20 times higher than the parent $V_{0.60}Ti_{0.40}$ alloy. Apart from this promising route for the enhancement of $J_c$ in the V-Ti alloys, the present study shows that in spite of the FM correlations introduced in the system due to the addition of Gd the superconductivity persists and in fact there is a marginal enhancement of the $T_{sc}$.